\documentclass[twocolumn,showpacs,floatfix,preprintnumbers,amsmath,amssymb]{revtex4}
\usepackage{graphicx}
\usepackage{epsfig}
\usepackage{dcolumn}
\usepackage{bm}

\begin{document}

\title{From energy-density functionals to mean field potentials: a systematic 
derivation}
\author{Ph. Chomaz$^{(1,2)}$ and K.H.O. Hasnaoui$^{(1,3)}$}
\affiliation{
(1) GANIL (DSM-CEA/IN2P3-CNRS), B.P.5027, F-14076 Caen c\'{e}dex 5, France \\
(2) CEA, Irfu/Dir, Centre de Saclay, F-91191 Gif-sur-Yvette, France\\
(3) Department of Physics, Tohoku University, Sendai 980-8578, Japan
 }


\begin{abstract}
In this paper we present a systematic method to solve the variational problem 
of the derivation of a self-consistent Kohn-Sham field from an arbitrary 
local energy functional. 
We illustrate this formalism with an application in nuclear physics and give the general 
mean field associated to the widely used Skyrme effective interaction.
\end{abstract}

\maketitle

\section{Introduction }

The density functional theory (DFT) is one of the most powerful theories
to deal with the intractable quantum many body problem for interacting systems 
with an arbitrary number of constituents\cite{kohn}. 
The original presentation of DFT from the Hohenberg-Kohn theorems\cite{hoh}
establishes an exact mapping between the total ground state energy 
and the local particle density $\rho (\mathbf{r})$ and has been essentially
used in electron structure theory. 
More general functionals of different local one-body observables can however
be used\cite{salsbury,stephens}. In particular in nuclear physics, the so called Hartree-Fock
formalism with effective density dependent interactions \cite{vautherin,schuck,bender}
can be seen as one of these extensions.

The practical implementation of density functional theory 
to electronic or nuclear structure problems
is done using self-consistent mean-field Kohn-Sham equations\cite{sham}. 
Within this framework, the interacting problem is reduced to a tractable 
problem of non-interacting particles moving in an effective potential. 
The functional form of this effective potential cannot be extracted from the bare 
interactions, and different approximations are done.
A simple local-density approximation (LDA) gives in general poor results, 
and more involved functionals including gradient terms, kinetic energy densities
and spin currents are currently employed\cite{GGA,chabanat}.  

For a given arbitrary functional the definition of the effective mean field 
potential is formally straightforward from the variational equation  
$\delta E=\mathrm{tr}\left( \hat{W}\delta \hat{\rho}\right)$. However the 
actual computation of $\hat{W}$ is tedious and complicated as soon as 
approximations going beyond LDA are considered.  
In this article, we present a systematic way to derive the mean field equation 
or equivalently the self-consistent Kohn-Sham
equations when a general functional of any local one-body observable is
considered. 

\section{Mean-field from density functional}

Let us consider the most general local energy-density, as a functional of all
possible local observables such as densities and currents. 

We first introduce the one body density as the observation 
associated to the one body density operator 
%of the $\hat{c}_{i}^{+}\hat{c}_{j}^{{}}$%
\begin{equation}
\rho _{ji}=<\hat{c}_{i}^{+}\hat{c}_{j}^{{}}>
\end{equation}
where $\hat{c}_{i}^{+}$ ($\hat{c}_{j}^{{}}$) creates (annihilates) a particle
in the single particle orbital  $\left| \varphi _{i}\right\rangle .$ The
formalism below can be easily extended to treat paring by introducing the
abnormal density $\kappa _{ji}=<\hat{c}_{i}^{+}\hat{c}_{j}^{+}>.$ 

All local observables can be expressed as a function of the one body density
 as local densities $\rho _{_{(A,B)}}(\mathbf{r})$
defined by the relation : 

\begin{equation}
\rho _{_{(A,B)}}(\mathbf{r})=\mathrm{tr}\left( \delta (\mathbf{\hat{r}}-%
\mathbf{r)}\hat{A}\hat{\rho}\hat{B}\right) 
\end{equation}

where $\mathbf{\hat{r}}$ is the position operator, $\hat{A}$ and $\hat{B}$
are one-body operators characterizing the considered density, and the
trace, $tr$, runs over all single particle states. 
In particular the choice $\hat{A}=\hat{B}=\hat{1}$ leads to the definition 
of the particle density, while $\hat{A}=\hat{B}=\hat{P}$ gives a kinetic energy 
density. More examples are worked out in the next section.

According to the most general version of DFT,
the local energy density  : 
\begin{equation}
E=\left\langle \psi \right| \hat{H}\left| \psi \right\rangle =\int \mathcal{H%
}(\mathbf{r})\mathrm{d}\mathbf{r}
\end{equation}
can be expressed as a functional of an ensemble of  local densities $\{\rho
_{_{A,B}}(\mathbf{r})\}$ associated to an ensemble of operators pairs $(\hat{A},\hat{B})$ such
as: 
\begin{equation}
\mathcal{H}(\mathbf{r})=\mathcal{H}[\rho _{_{(A,B)}}(\mathbf{r})]
\end{equation}

Using the variational derivation of mean field theory\cite{schuck},
the mean field hamiltonian is defined by the
equation 
\begin{equation}
\delta E=\mathrm{tr}\left( \hat{W}\delta \hat{\rho}\right)   \label{deltaE}
\end{equation}
We can thus write the
differential of the total energy in the form : 
\begin{eqnarray}
\delta E &=&\int \delta \mathcal{H}(\mathbf{r})\mathrm{d}\mathbf{r} 
\nonumber \\
&=&\sum_{(A,B)}\int \frac{\partial \mathcal{H}}{\partial \rho _{_{(A,B)}}}(%
\mathbf{r})\delta \rho _{_{(A,B)}}(\mathbf{r})\mathrm{d}\mathbf{r}  \nonumber
\\
&=&\sum_{(A,B)}\int \frac{\partial \mathcal{H}}{\partial \rho _{_{(A,B)}}}(%
\mathbf{r})\mathrm{tr}\left( \delta (\mathbf{\hat{r}}-\mathbf{r)}\hat{A}%
\delta \hat{\rho}\hat{B}\right) \mathrm{d}\mathbf{r} \nonumber\\
&=&\mathrm{tr}\left(\sum_{(A,B)}\frac{\partial \mathcal{H}}{\partial \rho
_{_{(A,B)}}}(\mathbf{\hat{r}})\hat{A}\delta \hat{\rho}\hat{B}\right)
\label{DeltaE2}
\end{eqnarray}
Identifying expression (\ref{DeltaE2}) with the mean-field hamiltonian
definition (\ref{deltaE}) we get : 
\begin{equation}
\hat{W}=\sum_{(A,B)}\hat{W}_{_{(A,B)}}=\sum_{(A,B)}\hat{B}\frac{\partial 
\mathcal{H}}{\partial \rho _{_{(A,B)}}}(\mathbf{\hat{r}})\hat{A}  \label{WAB}
\end{equation}

\subsection{Examples of usual operators}

In this part, we will identify the various terms, densities and currents,
which are used in the most common density dependent energy functionals with specific
local densities $\rho _{_{(A,B)}}(\mathbf{r})$

\subsubsection{Particle densities}

Let's start with the total particle density $\rho (\mathbf{r})$ that we can
write at first : 
\begin{equation}
\rho (\mathbf{r})=\sum_{\alpha }\left\langle \mathbf{r},\alpha \right| \hat{%
\rho}\left| \mathbf{r},\alpha \right\rangle 
\end{equation}
where the sum over $\alpha $ runs on all additional quantum numbers such as
spin and isospin. Introducing the operator $\delta \left( \mathbf{r}-\mathbf{%
\hat{r}}\right) $, we can rewrite this expression in a trace form: 
\begin{equation}
\rho (\mathbf{r})=\mathrm{tr}\left( \delta \left( \mathbf{r}-\mathbf{\hat{r}}%
\right) \hat{\rho}\right) 
\end{equation}
showing that $\hat{A}=\hat{B}=\hat{1}$ i.e. 
\begin{equation}
\rho (\mathbf{r})=\rho _{_{(1,1)}}(\mathbf{r})  \label{rho}
\end{equation}

Nuclear functionals generally use spin and isospin densities 
\begin{equation}
\rho _{ts}(\mathbf{r})=\rho _{_{(\tau_{t},\sigma_{s})}}(\mathbf{r})
\label{rhogene}
\end{equation}
where $\hat{\tau}_{t}$ is the $t$ component of the isospin Pauli operator
and $\hat{\sigma}_{s}$ the $s$ component of the spin Pauli matrix.
Of particular importance for asymmetric nuclear matter and exotic nuclei
is the isovector particle density 
\begin{equation}
\rho _{3}(\mathbf{r})=\rho _{_{(\tau_{3},1)}}(\mathbf{r})  \label{rho3}
\end{equation}
where $\hat{\tau}_{3}$ is the third component of the isospin operator.  

 Extending
the $\hat{\tau}$ and $\hat{\sigma}$ operators to include the identity $\tau
_{0}=\hat{1}$, and $\hat{\sigma}_{0}=\hat{1}$ the definition (\ref{rhogene})
includes the particle density (\ref{rho}) and the isospin density (\ref{rho3}%
), and can be taken as the most general. 
Note that since spin and isospin commute the order of the operators $\hat{%
T}$ and $\hat{\sigma}$ in (\ref{rhogene}) is not important, 
$\rho_{ts}=\rho_{st}$. It should also
be noticed that, in the nuclear case, 
the functional is often written in terms of proton and
neutron density. In such a case one should use the projector $\hat{\Pi }%
_{n}=(\hat{1}+\hat{\tau}_{3})/2$ for neutrons and $\hat{\Pi }_{p}=(\hat{1}-\hat{\tau}%
_{3})/2$ for protons.

\subsubsection{Kinetic densities}

The case of the  kinetic energy density $T(\mathbf{r})$ can be treated in
the same way :

\begin{equation}
T(\mathbf{r})=\sum_{\alpha }\sum_{k}\left\langle \mathbf{r},\alpha \right| \;%
\hat{p}_{_{k}}\hat{\rho}\;\hat{p}_{_{k}}\left| \mathbf{r},\alpha
\right\rangle 
\end{equation}
showing that $T(\mathbf{r})$ is the sum of 3 densities with  $(\hat{A},\hat{B%
})_{k}=(\hat{p}_{_{k}},\hat{p}_{_{k}}),$ $k=1,3$  i.e. 
\begin{equation}
T(\mathbf{r})=\sum_{k=1,3}\rho _{_{(p_{k},p_{k})}}(\mathbf{r})
\end{equation}
This definition can be extended to introduce spin and isospin as in (\ref
{rhogene}), in particular the isovector kinetic energy density 
\begin{equation}
T_{3}(\mathbf{r})=\sum_{k=1,3}\rho _{_{(\tau_{3}p_{k},p_{k})}}(\mathbf{r})
\end{equation}
More generally the spin and isospin kinetic energy densities can be defined
by
\begin{equation}
T_{kts}(\mathbf{r})=\rho _{_{(p_{k},p_{k}\tau_{t}\sigma_{s})}}(\mathbf{r})
\end{equation}

\smallskip 

\subsubsection{Spin currents}

Another important quantity in density functional theory is the spin current 
$\mathbf{J}(\mathbf{r})$ which is needed to describe the spin-orbit
interaction. It is defined by : 
\begin{equation}
\mathbf{J}(\mathbf{r})=-i\sum_{n,\alpha ,\alpha ^{\prime }}\left[ \mathbf{%
\nabla }_{\mathbf{r}}\varphi _{n}(\mathbf{r},\alpha )\right] \varphi
_{n}^{*}(\mathbf{r},\alpha ' )\times \left\langle \alpha ^{\prime }\right|
\sigma \left| \alpha \right\rangle +  c.c.
\end{equation}
where $\vec{\sigma}$ is the Pauli matrix. Such term can be easily derived
from the spin and isospin currents 
\begin{equation}
J_{kts}(\mathbf{r})=\rho _{_{(p_{k},\tau_{t}\sigma_{s})}}(\mathbf{r})+  c.c.
\end{equation}
using $J_{i}(\mathbf{r})=\sum_{ks=1\;}^{3}\varepsilon _{iks}\;J_{k0s}(%
\mathbf{r})$, where $\varepsilon $ is the anti-symmetric Levi-Cevita tensor.

\subsubsection{Gradient terms}

Energy functional are often written in terms of gradient of densities $\rho
_{_{(A,B)}}(\mathbf{r})$. Such terms can also be viewed as (combination of) 
generalized densities $\rho _{_{(A^{\prime },B^{\prime })}}(\mathbf{r})$ using the
relation 
\begin{eqnarray}
-i\mathbf{\nabla }\;\rho _{_{(A,B)}}(\mathbf{r}) &=&\mathrm{tr}\left( \delta (\mathbf{\hat{r}}-\mathbf{r})\left[\mathbf{\hat{p}},
\hat{A}\hat{\rho}\hat{B}\right]\right) \nonumber  \\
&=&\rho _{_{(\mathbf{p}A,B)}}(\mathbf{r})-\rho _{_{(A,B\mathbf{p}
)}}(\mathbf{r})
\end{eqnarray}

Higher order gradients can be similarly expressed if the above expressions are recursively
applied:

\begin{eqnarray}
\vec{\nabla}^{n}\rho
_{_{(A,B)}}(\mathbf{r})&=&(i)^{n}\sum_{j=0}^{n}(-1)^{n-j}C_{n}^{j} \cdot \nonumber \\
&\cdot& tr\left(\delta (\mathbf{\hat{r}}-\mathbf{r})
\hat{\mathbf{p}}^{j}
\hat{A}\hat{\rho}\hat{B}\hat{\mathbf{p}}^{n-j}
\right) \\
&=& \sum_{j=0}^{n}(i)^{n}(-1)^{n-j}C_{n}^{j}\rho
_{_{(\mathbf{p}^{j}A,B\mathbf{p}^{n-j})}}(\mathbf{r})\nonumber
\end{eqnarray}

where $C_{n}^{k}$ are binomial factors.

\section{APPLICATION TO SKYRME FUNCTIONALS}

As a practical application of the above formalism, let us consider the Skyrme
nuclear energy-density functional \cite{schuck,vautherin,chabanat} 
defined by : 

\begin{equation}
\mathcal{H}=\mathcal{K}+\mathcal{H}_{0}+\mathcal{H}_{3}+\mathcal{H}_{eff}+%
\mathcal{H}_{fin}+\mathcal{H}_{SO}+\mathcal{H}_{SG}
\end{equation}

where the $a_i$ coefficients are linear combination of the usual Skyrme 
parameters:

\begin{eqnarray*}
\mathcal{K} &=&\frac{T}{2m} \\
\mathcal{H}_{0} &=&a_{1}\rho ^{2}+a_{2}\left( \rho _{p}^{2}+\rho
_{n}^{2}\right)  \\
\mathcal{H}_{3} &=&\rho ^{\alpha }\left[ a_{3}\rho ^{2}+a_{4}\left( \rho
_{p}^{2}+\rho _{n}^{2}\right) \right]  \\
\mathcal{H}_{eff} &=&a_{5}T\rho +a_{6}\left[ T_{p}\rho _{p}+T_{n}\rho
_{n}\right]  \\
\mathcal{H}_{fin} &=&a_{7}(\mathbf{\nabla }\rho )^{2}+a_{8}\left[ (\mathbf{%
\nabla }\rho _{p})^{2}+(\mathbf{\nabla }\rho _{n})^{2}\right]  \\
\mathcal{H}_{so} &=&a_{9}\left[ \mathbf{J}.\mathbf{\nabla }\rho +\mathbf{J}%
_{p}.\mathbf{\nabla }\rho _{p}+\mathbf{J}_{n}.\mathbf{\nabla }\rho
_{n}\right]  \\
\mathcal{H}_{sg} &=&a_{10}\mathbf{J}^{2}+a_{11}\left[ \mathbf{J}_{p}^{2}+%
\mathbf{J}_{n}^{2}\right] 
\end{eqnarray*}
with 
\begin{eqnarray*}
a_{1} &=&\frac{t_{0}}{4}\left( 2+x_{0}\right) ,\;a_{2}=-\frac{t_{0}}{4}%
\left( 2x_{0}+1\right)  \\
a_{3} &=&\frac{t_{3}}{24}\left( 2+x_{3}\right) ,\;a_{4}=-\frac{t_{3}}{24}%
\left( 2x_{3}+1\right)  \\
a_{5} &=&\frac{1}{8}\left[ t_{1}\left( 2+x_{1}\right) +t_{2}\left(
2+x_{2}\right) \right]  \\
a_{6} &=&\frac{1}{8}\left[ t_{2}\left( 2x_{2}+1\right) -t_{1}\left(
2x_{1}+1\right) \right]  \\
a_{7} &=&\frac{1}{32}\left[ 3t_{1}\left( 2+x_{1}\right) -t_{2}\left(
2+x_{2}\right) \right]  \\
a_{8} &=&-\frac{1}{32}\left[ t_{2}\left( 2x_{2}+1\right) +3t_{1}\left(
2x_{1}+1\right) \right]  \\
a_{9} &=&\frac{W_{0}}{2}\;,\;\;a_{10}=-\frac{1}{16}\left(
t_{1}x_{1}+t_{2}x_{2}\right)  \\
a_{11} &=&\frac{1}{16}\left( t_{1}-t_{2}\right) 
\end{eqnarray*}

We first write the various terms in terms of generalized
densities $\rho _{_{({A},{B})}}$%

\begin{eqnarray*}
\mathcal{K} &=&\frac{\sum_{k=1}^{3}\rho _{_{({p}_{k},{p}_{k})}}}{2m} \\
\mathcal{H}_{0} &=&a_{1}\rho _{_{({1},{1})}}^{2}+a_{2}(\rho _{_{({\Pi} %
_{n},{1})}}^{2}+\rho _{_{({\Pi} _{p},{1})}}^{2}) \\
\mathcal{H}_{3} &=&a_{3}\rho _{_{({1},{1})}}^{2+\alpha }+a_{4}\rho
_{_{({1},{1})}}^{\alpha }(\rho _{_{({\Pi} _{n},{1})}}^{2}+\rho _{_{({\Pi} _{p},{1})}}^{2})
\\
\mathcal{H}_{eff} &=&a_{5}\sum_{k=1}^{3}\rho _{_{({p}_{k},{p}_{k})}}\rho _{_{({1},{1})}}
\\
&+&a_{6}(\rho _{_{({\Pi} _{n}{p}_{k},{p}_{k})}}\rho _{_{({\Pi} _{n},{1})}}+\rho _{_{(%
{\Pi} _{p}{p}_{k},{p}_{k})}}\rho _{_{({\Pi} _{p},{1})}}) \\
\mathcal{H}_{fin} &=&-a_{7}\sum_{k=1}^{3}(\rho _{_{({p}_{k},{1})}}-\rho
_{_{({1},{p}_{k})}})^{2}  \\
&+&a_{8}\left[ (\rho _{_{({p}_{k},{\Pi} _{n})}}-\rho _{_{({\Pi} %
_{n},{p}_{k})}})^{2}+(\rho _{_{({p}_{k},{\Pi} _{p})}}-\rho _{_{({\Pi} %
_{p},{p}_{k})}})^{2}\right]  \\
\mathcal{H}_{so} &=&i a_{9}\sum_{iks=1\;}^{3}\varepsilon _{iks}[\rho
_{_{({p}_{k},{\sigma} _{s})}}.(\rho _{_{({p}_{i},{1})}}-\rho _{_{({1},{p}_{i})}}) \\
&+&\rho _{_{({\Pi} _{p}{p}_{k},{\sigma} _{s})}}.(\rho _{_{({p}_{i},{\Pi} _{p})}}-\rho
_{_{({\Pi} _{p},{p}_{i})}}) \\
&+&\rho _{_{({\Pi} _{n}{p}_{k},{\sigma} _{s})}}.(\rho _{_{({p}_{i},{\Pi} _{n})}}-\rho
_{_{({\Pi} _{n},{p}_{i})}})] \\
\mathcal{H}_{sg} &=&a_{10}^{{}}\sum_{iksk^{\prime }s^{\prime
}=1\;}^{3}\varepsilon _{iks}\varepsilon _{ik^{\prime }s^{\prime }}\rho
_{_{({p}_{k},{\sigma} _{s})}}\rho _{_{({p}_{k^{\prime }},{\sigma} _{s^{\prime }})}}
\\
&+&a_{11}\sum_{iksk^{\prime }s^{\prime }=1\;}^{3}\varepsilon
_{iks}\varepsilon _{ik^{\prime }s^{\prime }}(\rho _{_{({p}_{k},{\Pi} _{p}{\sigma}
_{s})}}\rho _{_{({p}_{k^{\prime }},{\Pi} _{p}{\sigma} _{s^{\prime }})}} \\
&+&\rho _{_{({p}_{k},{\Pi} _{n}{\sigma} _{s})}}\rho _{_{({p}_{k^{\prime }},{\Pi} %
_{n}{\sigma} _{s^{\prime }})}})
\end{eqnarray*}

Using relation (\ref{WAB}), the involved functional derivatives of eq.(\ref{deltaE}) are 
reduced to simple derivatives of scalar expressions:

\begin{eqnarray*}
W_{\mathcal{K}} &=&\frac{\mathbf{\hat{p}}^{2}}{2m} \\
W_{\mathcal{H}_{0}} &=&2a_{1}\rho (\mathbf{\hat{r})}+2a_{2}(\hat{\Pi }%
_{n}\rho _{n}(\mathbf{\hat{r})}+\hat{\Pi }_{p}\rho _{p}(\mathbf{\hat{r})}) \\
W_{\mathcal{H}_{3}} &=&(2+\alpha )a_{3}\rho ^{1+\alpha }(\mathbf{\hat{r})} \\
&+&\alpha a_{4}\rho ^{\alpha -1}(\mathbf{\hat{r})}(\rho _{n}^{2}(\mathbf{%
\hat{r})}+\rho _{_{p}}^{2}(\mathbf{\hat{r})}) \\
&+&2a_{4}\rho ^{\alpha -1}(\mathbf{\hat{r})}(\hat{\Pi }_{n}\rho _{n}(\mathbf{%
\hat{r})}+\hat{\Pi }_{p}\rho _{p}(\mathbf{\hat{r})}) \\
W_{\mathcal{H}_{eff}} &=&\sum_{k=1}^{3}a_{5}\hat{p}_{k}\rho (\mathbf{\hat{r})}%
\hat{p}_{k}+a_{5}T(\mathbf{\hat{r})} \\
&+&\sum_{k=1}^{3}a_{6}\hat{\Pi }_{n}\hat{p}_{k}\rho _{n}(\mathbf{\hat{r})}\hat{%
p}_{k}+a_{6}\hat{\Pi }_{n}T_{n}(\mathbf{\hat{r})} \\
&+&\sum_{k=1}^{3}a_{6}\hat{\Pi }_{p}\hat{p}_{k}\rho _{p}(\mathbf{\hat{r})}\hat{%
p}_{k}+a_{6}\hat{\Pi }_{p}T_{p}(\mathbf{\hat{r})} \\
W_{\mathcal{H}_{fin}} &=&-2a_{7}\Delta \rho (\mathbf{\hat{r})} \\
&-&2a_{8}(\hat{\Pi }_{n}\Delta \rho _{n}(\mathbf{\hat{r})}+\hat{\Pi }_{p}%
\Delta \rho _{p}(\mathbf{\hat{r}))} \\
W_{\mathcal{H}_{so}} &=&a_{9}(\mathbf{\nabla }\rho (\mathbf{\hat{r}) 
+\nabla J}(\mathbf{\hat{r})} \\
&+&\hat{\Pi }_{n}\mathbf{\nabla }\rho _{n}(\mathbf{\hat{r}) }%
+\hat{\Pi }_{n}\mathbf{\nabla J}_{n}(\mathbf{\hat{r})} \\
&+&\hat{\Pi }_{p}\mathbf{\nabla }\rho _{p}(\mathbf{\hat{r}) }%
+\hat{\Pi }_{p}\mathbf{\nabla J}_{p}(\mathbf{\hat{r}))} \\
W_{\mathcal{H}_{sg}} &=&2a_{10}\mathbf{J}(\mathbf{\hat{r}) } \\
&+&2a_{11}(\hat{\Pi }_{n}\mathbf{J}_{n}(\mathbf{\hat{r}) +}\hat{%
\Pi }_{p}\mathbf{J}_{p}(\mathbf{\hat{r}) )}
\end{eqnarray*}

The one body mean field operator for the charge $q=n,p$
can be expressed in a more compact form as :
 
\begin{equation}
\hat{W}_{q}=\mathbf{\hat{p}}\frac{1}{2m(\mathbf{\hat{r}})}\mathbf{\hat{p}}%
+U_{q}(\mathbf{\hat{r}})+\left(\mathbf{\hat{p}}\mathbf{V}_{q}(\mathbf{\hat{r}})+\mathbf{V}_{q}(\mathbf{\hat{r}})\mathbf{\hat{p}}\right)\times \mathbf{\hat{\sigma}}%
\end{equation}
where the effective mass is 
\begin{equation}
\frac{1}{2m(\mathbf{\hat{r}})}=\left[ \frac{1}{2m}+a_{5}\rho (\mathbf{\hat{r}%
})+a_{6}\rho _{q}(\mathbf{\hat{r}})\right] 
\end{equation}
while the central potential reads 
\begin{eqnarray}
U_{q}(\mathbf{\hat{r}}) &=&2a_{1}\rho (\mathbf{\hat{r}})+2a_{2}\rho _{q}(%
\mathbf{\hat{r}})+a_{3}\left( 2+\alpha \right) \rho ^{1+\alpha }(\mathbf{%
\hat{r}})  \nonumber \\
&+&a_{4}\alpha \rho ^{\alpha -1}(\mathbf{\hat{r}})(\rho _{n}^{2}(\mathbf{%
\hat{r}})+\rho _{p}^{2}(\mathbf{\hat{r}}))+2a_{4}\rho _{q}(\mathbf{\hat{r}}%
)\rho ^{\alpha }(\mathbf{\hat{r}}) \\
&+&a_{5}T(\mathbf{\hat{r}})+a_{6}T_{q}(\mathbf{\hat{r}})  \nonumber \\
&-&2a_{7}\Delta \rho (\mathbf{\hat{r}})-2a_{8}\Delta \rho _{q}(\mathbf{\hat{r%
}})+a_{9}\left( \mathbf{\nabla J}(\mathbf{\hat{r})+\nabla J}_{q}(\mathbf{%
\hat{r})}\right)   \nonumber
\end{eqnarray}
and the spin orbit part is given by 
\begin{equation}
\mathbf{V}_{q}(\mathbf{r})=a_{9}\left( \mathbf{\nabla }\rho (\mathbf{\hat{r})}+%
\mathbf{\nabla }\rho _{q}(\mathbf{\hat{r})}\right) +2a_{10}\mathbf{J}(%
\mathbf{\hat{r})}+2a_{11}\mathbf{J}_{q}(\mathbf{\hat{r})}
\end{equation}

consistently with the expressions given in the literature.  

\section{CONCLUSION}

In this paper we have introduced a simple and powerful method to solve 
the variational problem of the derivation of a mean field potential from 
and arbitrary energy-density functional. Observing that the most general DFT energy functional  can be expressed in 
terms of generalized local one-body densities  $\rho
_{_{(A,B)}}(\mathbf{r})=\mathrm{tr}\left( \delta (\mathbf{\hat{r}}-\mathbf{r)%
}\hat{A}\hat{\rho}\hat{B}\right) $, the functional derivative defining the 
Kohn-Sham or HF field $\hat{W}=\delta{E}/\delta\hat{\rho}$ can be reduced
to the calculation of ordinary derivatives $\partial {\cal{H}} /\partial \rho_{(A,B)}$
of the associated energy density. As an example we have worked out the well known expression of the 
general Skyrme mean-field. This method gives a simple technique 
to take into account new terms in future more sophisticated generalized
functionals. The autors want to thank M. Bender for the useful discussions.


\begin{thebibliography}{99}
\bibitem{gross} R. Dreizler, E. Gross, \textit{Density Functional Theory}, 
                                 Plenum Press, New York (1995).

\bibitem{kohn} { W. Kohn, Rev. Mod. Phys., Vol. 71, No. 5, (1999) and
refs. therein. }

\bibitem{hoh} P. Hohenberg and W. Kohn, Phys. Rev. B136 (1964) 864

\bibitem{salsbury} F.R. Salsbury, C.J. Grayce and R.A. Harris,  Phys. Rev. A 50, (1994) 3089. 
%�Magnetic-field density-functional theory,�

\bibitem{stephens}P. J. Stephens, F. J. Devlin, C. F. Chabalowski, and M. J. Frisch, J. Phys. Chem. 98 (1994) 11623.
%DFT+HF exchange

\bibitem{vautherin}  {D. Vautherin, D.M. Brink, Phys.Rev. C5(1972) 626.}

\bibitem{schuck}  {P. Ring, P. Schuck \textit{The nuclear Many-Body Problem}%
, Springer-Verlag Berlin (1980)}


\bibitem{bender}   M. Bender, P.H. Heenen, and P.-G. Reinhard, Rev. Mod. Phys. 75 (2003) 121-180.

%\bibitem{meyer} {J. Meyer \textit{Interactions effectives, th\'eories de champ moyen, masses et rayons nucl\'eaires}Annales de physique.Volume 28.N3.2003.}

\bibitem{sham} W. Kohn and L. J. Sham, Phys. Rev. 140 (1965) A1133

%\bibitem{GGA}D. C. Langreth and J. P. Perdew, Phys. Rev. B 21, 5469(1980).
%generalized gradient approximation - original

\bibitem{GGA}Y. Zhang and W. Yang, Phys. Rev. Lett. 80, 890 (1998).

\bibitem{chabanat} E.Chabanat, P.Bonche, P.Haensel, J.Meyer, R.Schaeffer,
Nucl.Phys. A635, 231 (1998).
 
%\bibitem{Almbladh} { Almbladh, C. O., and U. von Barth, 1985, Phys. Rev. B
%31, 3231.}

%\bibitem{Becke}{  Becke, A. D., 1996, J. Chem. Phys. 104, 1040.}

\end{thebibliography}
\end{document}